\documentclass[conference]{IEEEtran}
\usepackage{amsmath}
\usepackage{amssymb}                                           
\usepackage{MnSymbol}
\usepackage{amsxtra}
\usepackage{bbold}
\usepackage{mathdots}
\usepackage{mathtools}
\usepackage{pdfcolmk}
\usepackage{graphicx}                                        
\usepackage{algorithm}
\usepackage{algorithmic} 
\usepackage{mathrsfs}
\usepackage{todonotes}

\usepackage[utf8]{inputenc}                                         
\usepackage[T1]{fontenc}
\usepackage[pdfborder={0 0 0},hypertexnames=false,extension=pdf]{hyperref} 
\usepackage{verbatim}

\usepackage{srcltx}
\usepackage[english]{babel}

\usepackage[shortlabels]{enumitem}
\usepackage[all]{xy}

\usepackage[style=ieee,    
            firstinits=true, 
            url=false,
            backend=bibtex,
            maxbibnames=99,
            isbn=false,
            natbib=true,
            hyperref=true,
            bibencoding=utf8]{biblatex}
\AtEveryBibitem{\clearfield{note}}   
\DefineBibliographyStrings{english}{%
  references = {},
}

\bibliography{refs,jabref_philipp_utf2}

\newcommand{\norm}[1]{\left\|#1\right\|}

\newcommand{\abs}[1]{\left|#1\right|}

\newcommand{\diag}{\operatorname{diag}}

\let\forallalt\forall
\renewcommand{\forall}{\;\forallalt\;}

\let\refalt\ref
\renewcommand{\ref}[1]{(\refalt{#1})}

\renewcommand{\vec}{\mathbf}

\newcommand{\alp}{\ensuremath{\alpha}}

\newcommand{\lam}{\ensuremath{\lambda}}

\newcommand{\vLam}{\ensuremath{\mathbf{\Lambda}}}

\newcommand{\eps}{\ensuremath{\epsilon}}

\newcommand{\Pset}{{\ensuremath{\mathcal{P}}}}

\newcommand{\Omi}{\ensuremath{\mathcal{O}}}

\newcommand{\ome}{\ensuremath{\omega}}

\newcommand{\vOme}{\ensuremath{\mathbf{\Omega}}}

\newcommand{\vta}{\ensuremath{\tilde{\mathbf a}}}

\newcommand{\hvu}{\ensuremath{\hat{\vu}}}
\newcommand{\hvn}{\ensuremath{\hat{\vn}}}

\newcommand{\A}{{\ensuremath{\mathcal{A}}}}

\newcommand{\Symnull}{{\ensuremath{\mathcal{S}^{\circ}_0}}}

\newcommand{\SymK}{{\ensuremath{\mathcal{{S}}^{\circ}_{\!K}}}}
\newcommand{\SymtK}{{\ensuremath{\mathcal{{S}}^{\circ}_{\!K'}}}}

\newcommand{\C}{{\ensuremath{\mathbb C}}}

\newcommand{\R}{{\ensuremath{\mathbb R}}}

\newcommand{\N}{{\ensuremath{\mathbb N}}}

\newcommand{\Fmatrix}{{\ensuremath{\mathbf F}}}

\newcommand{\FmatrixN}{{\ensuremath{\mathbf F}_{\!N}}}
\newcommand{\FmatrixP}{{\ensuremath{\mathbf F}_{\!P}}}

\newcommand{\id}{{\ensuremath{\mathbb 1}}}

\newcommand{\va}{{\ensuremath{\mathbf a}}}
\newcommand{\vb}{{\ensuremath{\mathbf b}}}
\newcommand{\vc}{{\ensuremath{\mathbf c}}}
\newcommand{\vd}{{\ensuremath{\mathbf d}}}
\newcommand{\ve}{{\ensuremath{\mathbf e}}}
\newcommand{\vg}{{\ensuremath{\mathbf g}}}

\newcommand{\vh}{\ensuremath{\mathbf h }}                         
\newcommand{\vr}{\ensuremath{\mathbf r}}                         
\newcommand{\vn}{\ensuremath{\mathbf n}}                         
\newcommand{\vs}{{\ensuremath{\mathbf s}}}
\newcommand{\vu}{{\ensuremath{\mathbf u}}}

\newcommand{\vx}{{\ensuremath{\mathbf x}}}

\newcommand{\vz}{{\ensuremath{\mathbf z}}}

\newcommand{\vth}{\ensuremath{\tilde{\mathbf h}}}

\newcommand{\ihve}{{\ensuremath{\check{\mathbf e}}}}

\newcommand{\hvs}{{\ensuremath{\hat{\mathbf s}}}}
\newcommand{\hvx}{{\ensuremath{\hat{\mathbf x}}}}
\newcommand{\hvy}{{\ensuremath{\hat{\mathbf y}}}}

\newcommand{\vA}{{\ensuremath{\mathbf A}}}

\newcommand{\vG}{{\ensuremath{\mathbf G}}}
\newcommand{\vtG}{{\ensuremath{\tilde{\mathbf G}}}}

\newcommand{\valp}{\ensuremath{\boldsymbol{ \alp}}}

\newcommand{\zero}{{\ensuremath{\mathbf 0}}}

\newcommand{\vS}{\ensuremath{\mathbf S }}                         

\newcommand\rk{\operatorname{rk}} 
\definecolor{gray}{rgb}{0.3,0.3,0.3}
{\color{black}}                      
{\color{black}}

{\color{black}}                      
{\color{black}}

\newcommand{\figref}[1]{Figure~\ref{#1}}

\newcommand{\noi}{\noindent}

\ifx\definition\undefined
\newtheorem{definition}{Definition}         
\fi
\ifx \@definition \@empty
\newtheorem{definition}[theorem]{Definition}   
\fi
\ifx\corrolary\undefined
\newtheorem{corrolary}{Corrolary} 
\fi
\ifx\conjecture\undefined
\newtheorem{conjecture}{Conjecture}         
\fi
\ifx\theorem\undefined
\newtheorem{theorem}{Theorem}         
\fi
\ifx\lemma\undefined
\newtheorem{lemma}{Lemma}
\fi
\ifx\question\undefined
\newtheorem{question}{\color{red}{Question}}         
\fi

{\par\noindent{\em Beweis\/}.}%
{\hspace*{\fill}{\qed}\vspace{1ex}\par}
{\par\noindent{\em Proof\/}.}%
{\par}

{\hspace*{\fill}{\lightning}\vspace{1ex}\par}
{\par\vspace{1.5ex}\noindent{\em Remark\/}.}
{\par\vspace{1.5ex}}
\ifx\remark\undefined
\newenvironment{remark}%
{\par\vspace{1.5ex}\noindent{\em Remark\/}.}
{\par\vspace{1.5ex}}
{\par\vspace{1.5ex}\noindent{\em Recall\/}.}
{\par\vspace{1.5ex}}
{\par\vspace{1.5ex}\noindent{\em Example\/}. }
{\par\vspace{1.5ex}}
{\noi\vspace{0.5ex}\small}
{\vspace{0.5ex}\par\normalsize}

\newcounter{Examplecount}
{\renewcommand{\labelenumi}{(\roman{enumi})}\begin{list}{\labelenumi}
{\usecounter{enumi}\setlength{\labelwidth}{1.5cm}\setlength{\topsep}{0.3cm}\setlength{\itemsep}{-3pt}}}
{\end{list}}
{\renewcommand{\labelenumi}{(\arabic{enumi})}\begin{list}{\labelenumi}
{\usecounter{enumi}\setlength{\labelwidth}{1.5cm}\setlength{\topsep}{0.3cm}\setlength{\itemsep}{-3pt}}}
{\end{list}}
{\renewcommand{\labelenumi}{$\bullet$}\begin{list}{\labelenumi}
{\setlength{\labelwidth}{1.5cm}\setlength{\topsep}{0.3cm}\setlength{\itemsep}{-2pt}}}
{\end{list}}

\makeatletter 
\newcommand{\sprod}[2]{\ensuremath{%
\setbox0=\hbox{\ensuremath{#2}}
\dimen@\ht0
\advance\dimen@ by \dp0
\left(\left.#1\rule[-\dp0]{0pt}{\dimen@}\right|#2\right)}}
\makeatother 

\makeatletter
\newcommand{\set}[2]{\ensuremath{%
\setbox0=\hbox{\ensuremath{#2}}
\dimen@\ht0
\advance\dimen@ by \dp0
\left\{\left.#1\rule[-\dp0]{0pt}{\dimen@}\;\right|\;#2\right\} }}
\makeatother

\newcommand\trace{\operatorname{tr}} 
 
\newcommand\tr{\operatorname{tr}}

\newcommand{\argmin}[1]{\underset{#1}{\operatorname{argmin}}}

\newcommand{\Norm}[1]{\ensuremath{ \left\|#1\right\| }}

\newcommand{\cc}[1]{{\ensuremath{\overline{#1}}}} 

\DeclareMathOperator{\supp}{supp}
\renewcommand{\argmin}{\operatornamewithlimits{argmin}}

\newcommand{\vy}{\ensuremath{\mathbf y}}

\newcommand{\vQ}{\ensuremath{\mathbf Q}}
\newcommand{\vR}{\ensuremath{\mathbf R}}

\ifx \@paragraph \@empty
\makeatletter
\renewcommand\paragraph{\@startsection
{paragraph}{4}{\z@}{-3.5ex plus-1ex minus-.2ex}%
{1.3ex plus.2ex}{\normalfont\itshape}}
\makeatother
\fi

\makeatletter
\providecommand\@dotsep{5}

\makeatother

\newcommand{\CP}[1]{\ensuremath{ #1_{_{\text{CP}}}}}

\renewcommand{\Re}{\ensuremath{\operatorname{Re}}}
\renewcommand{\Im}{\ensuremath{\operatorname{Im}}}

\newcommand{\vhtr}{\overline{\vh}_-}

\begin{document}
%
\title{OFDM Channel Estimation via Phase Retrieval}

\author{\IEEEauthorblockN{Philipp Walk}
\IEEEauthorblockA{Department of Electrical Engineering\\
California Institute of Technology\\
MC 136-93, Pasadena, CA 91125\\
Email: pwalk@caltech.edu}
\and
\IEEEauthorblockN{Henning Becker}
\IEEEauthorblockA{Theoretical Information Technology\\
Technical University Munich\\
Theresienstrasse 90, 80290 M{\"u}nchen\\
Email: henning.becker@tum.de}
\!\and \!
\IEEEauthorblockN{Peter Jung}
\IEEEauthorblockA{
 Communications \& Information Theory\\ 
  Technical University Berlin (TUB)\\
  Einsteinufer 25, 10587 Berlin\\
  Email: peter.jung@tu-berlin.de}
}

\maketitle

\begin{abstract}
  Pilot--aided channel estimation is nowadays a standard component in
  each wireless receiver enabling coherent transmission
  of complex--valued constellations, only affected by noise and
  interference. Whenever these disturbances are sufficiently small
  and long data frames are used,
  high data rates can be achieved and the resource overhead due to the
  pilots vanishes asymptotically. On the other, it is expected that
  for the next generation of mobile networks not only data rate is in the
  main focus but also low latency, short and sporadic messages, 
  massive connectivity, distributed\&adhoc processing and robustness with respect to asynchronism.
  Therefore a review of several well-established principles in
  communication has been started already.
  
  A particular implication when using complex--valued pilots
  is that these values have to be \emph{known at the 
  receiver} and therefore these resources can not be used
  simultaneously for user data. For an OFDM-like multicarrier scheme
  this means that pilot tones (usually placed
  equidistantly according to the Nyquist theorem) are allocated
  with globally known amplitudes and phases to reconstruct the channel impulse
  response. Phases are designed and allocated globally which is in contrast
  to a distributed infrastructure.

  In this work we present therefore a new \emph{phaseless pilot} scheme where
  only pilot amplitudes need to be known at the receiver, i.e., phases are available
  again and can be used for various other purposes. The idea is based on a
  phase retrieval result for symmetrized and zero-padded magnitude
  Fourier measurements obtained by two of the authors. 
  The phases on the pilot tones can now be used to carry additional
  user--specific data or  compensate for other signal
  characteristics, like the PAPR. 
\end{abstract}

\section{Introduction}

OFDM is a well--established common multicarrier modulation scheme for high--rate data transmission in time--invariant
channels. It is used for example in IEEE-802.11a/g, LTE and DSL. New pulse--shaped
multicarrier system which almost diagonalize typical mobile channels
and are robust in the presence of asynchronism are proposed as candidates for the next generation of
mobile systems. Most of these schemes use time--frequency multiplexing
and are in this direction similar to OFDM.
To obtain information about the channel for each subchannel pilot-based methods are used in most
cases.  Hence some resources are spent solely for channel estimation
and this decreases the overall spectral efficiency. Obviously, whenever noise and
interference contributions are sufficiently small, the channel does
not change too rapidly, latency and energy is not an issue,
high data rates can be achieved in this way. Asymptotically this
overhead can then be neglected. On the other hand, the next
generation of wireless mobile communication networks are much more
than providing high data rates. It is expected that with the next cycles
completely different communication scenarios will be in the focus
where low latency, sporadic and short message support,
asynchronous and non--orthogonal operation modes and distributed/adhoc
and local processing are the major challenges.

Under these prerequisites it is important again to review several
well--established principles of communication including also the
tradeoff between pilot--overhead and amount of original information transfer.
To come up with efficient blind estimation techniques which are able
to operate on a short data frame basis is challenging. Intuitively it
is clear that further apriori assumptions on the channel characteristics,
like certain compressibility properties of the channel, for example
sparsity, have to be exploited. From practical side, it is also difficult
to implement such approaches efficiently in near future.  
However, a first step hereby would be the use of pilot phases for
different tasks, i.e., to estimate the channel coefficients only by 
the magnitudes on the received pilot tones. 
The pilot phases are the available again for carring user--specific
data. There are endless potential applications: (i) use pilot phases
as a second data layer; (ii) use phase--shift keying
alphabets (with constant and known amplitude) without specific pilot
tones; (iii) use pilot phases for data--dependent signal compensation 
(like tone--reservation for reducing the peak--to--average power ratio) and (iv) phase--preprocessing to 
avoid data--pilot interference in Offset-QAM--based multicarrier modulation.
In addition, a central pilot--phase management is completely avoided, 
which has advantages in dense network structures with massive access.
On the other hand, these benefits come not for free. Since
conventional pilot--aided channel estimation is essentially the 
inversion of a linear problem it is in most cases stable, either 
by considering pilot placement under the Nyquist criterion or
exploiting Compresses Sensing for recovery under sparsity priors.
Phaseless recovery in turn is a quadratic problem and achieving
stability in this particular setting is much more challenging.

First work in this
direction has already be shown by the authors in \cite{WBJ15a}. 
The intention of this contribution is (i) to present our ideas in a
more clear fashion; (ii) to further explain our scheme in the setting
of phase retrieval and (iii) to discuss also differences to phase retrieval
and finally to show some performance results achieved by
an iterative estimation algorithm.
The structure of the paper is as follows: First we give background
information for the phase retrieval problem with further details 
for Fourier measurements. We explain some differences to our setup
of symmetrized Fourier measurements. Using these investigations
we propose a particular frame structure to support phaseless pilots
for a semi--blind channel estimation (using only magnitudes). We will
establish an iterative two--stage algorithm which can recover the
channel impulse response up to global sign from
magnitudes on the pilot tones. We will show simulations
demonstrating that performance in terms of error rates.
\\
{\em Notation:} We denote by capital letters integers and write for the set of the first $N$ integers
$[N]=\{0,1,\dots,N-1\}$. Bold letters denote  vectors and bold capital letters refer to matrices.  We denote by
$\FmatrixN$ the unitary
$N-$dimensional discrete Fourier transform (DFT)  giving  element-wise for $k,l\in[N]$ as
$(\FmatrixN)_{lk}=\ome_N^{lk}/\sqrt{N}$ where $\ome_N=e^{-i2\pi/N}$ is the $N$th root of unity. We will abbreviate
 $\Fmatrix\vx=\hvx=\FmatrixN\vx$ whenever the dimension of $\vx\in\C^N$ is clear from context.   The pointwise product
is denoted by $\odot$ and the linear convolution by $*$ where $\circledast$ refers to the circular convolution. We use
further $\cc{\vx}$ to denote complex conjugation of all vector coefficients and $\vR\vx=\vx_-=(x_{N-1},\dots, x_0)^T$ for the
time reversal of $\vx\in\C^N$. The  $N\times N$ circular (down) shift matrix and time-reversal matrix is given by
\begin{align}
  \vS =\vS_N =\left(\begin{smallmatrix}
      0  & \cdots & 0 & 1 \\
      1 &  \cdots& 0 & 0\\
      \vdots &  \ddots & & \vdots \\
      0 &  \cdots &1& 0 \\ 
    \end{smallmatrix}\right)      \quad,\quad\vR =\vR_N =\left(\begin{smallmatrix}
      0  & \cdots & 0 & 1 \\
      0 &  \cdots& 1 & 0\\
      \vdots &  \iddots & & \vdots \\
      1 &  \cdots &0& 0 \\ 
    \end{smallmatrix}\right) 
  \label{eq:notation:shift:treverse}
\end{align}
We set $\ve_0=(1,0,\dots,0)^T$ such that the circular shifts $\vS\ve_0=\ve_k$ define the Euclidean basis in $N-$dimension  $\{\ve_k\}_{k=0}^{N-1}$.
%

\section{Background}

\subsection{Background on Phase Retrieval}

The recovery of a signal from magnitude measurements is known as the
\emph{phase retrieval problem}. It has a long history beginning with
the work of Gerchberg and Saxton in the $70$'s \cite{GS72}. Later,
Fienup \cite{Fie78b} also considered this problem and gave explicit
reconstruction algorithms for the phase from magnitude 
of Fourier measurements. 
One of the challenging tasks in phase retrieval is to determine the necessary and sufficient number of
measurements for ensuring injectivity or even stability.  
Since one cannot distinguish between numbers of unit modulus from the magnitude of a linear measurement,
such statements can hold here only up to a global phase.

\paragraph{Generic Measurements} Candes et.al.  \cite{CSV12} have
shown stable recovery w.h.p. of
$N-$dimensional complex-valued signals from the magnitude of
$\Omi(N\log N)$ Gaussian measurements. A
more principal result from Balan et al. in \cite{BCE06} shows
that injectivity holds for $4N-2$  generic measurements. Moreover, they could give a fast reconstruction algorithm in \cite{BBCE07}. Using projection
methods, Mondragon and Voroninski could even show in \cite{MV13} injectivity from $4N-3$ generic
measurements.  
However, a practical construction and implementation of measurements at this limiting number
seems to be rather hard, but it serves as an ultimate theoretical bound.

\paragraph{Fourier Measurements}
It is well--known that for Fourier measurements further ambiguities,
like for example conjugation, translation and reflection, can not be
avoided. Further non--trivial ambiguities are characterized in
\cite{BP15}. Therefore, several modifications of ``pure'' Fourier
measurements are proposed, depending on the application.
For example,  non-linear or interference--based  approaches are considered to provide unique phase reconstruction.
Wang \cite{Wan13} presented a method where interference with a known signal $\vy\in\C^N$ helps to
recover a signal $\vx\in\C^N$ up to a global sign from only $3N$ Fourier measurements $|\Fmatrix(\vx+\ome\vy)|^2$ where
$\ome\in\C$ is a root of unity.  
To mitigate the ambiguities in phase retrieval the early approaches
also try to explore sparsity properties on the signal.
But it shows up that either oversampling nor sparsity helps to dissolve all ambiguities in the one-dimensional phase
retrieval.  Lu and Vetterli also use sparsity for spectral factorization of real valued impulse
responses  \cite{LV11}. Moreover, they also give a reconstruction algorithm.

\paragraph{Sparse Phase Retrieval} With the advent of compressed sensing, generalized (generic) phase retrieval with
sparsity priors came back in the research focus.  
For \emph{real} $S$--sparse signals, Eldar and Mendelson \cite{EM12}
established  stable recovery w.h.p. from
$\Omi(S\log (eN/S))$ subgaussian measurements with high probability.
The result \cite{EFS13} of Ehler, Fornasier and Sigl extends this to
the complex case and they provide an explicit reconstruction
algorithm, but the signal needs to have strong decay properties.  Wang and Xu
\cite{WX13} stated injectivity for $S-$sparse complex-valued signals from $4S-2$ generic measurements as long as $S<N$.

\subsection{Fourier Phase Retrieval and Conjugate-Symmetry}

To dissolve the inherent ambiguities in the phase retrieval problem
with Fourier measurements, as (circular) time-reversal, time-shift,
and a global phase factor, see e.g. \cite{BP15} and \cite{JOH13}, it
is also possible to 
consider or even construct (depending on the application) a further structure
on the signal itself, which excludes the ambiguities right away, up to global sign. This structure is indeed known.
Whenever the signal is invariant under circular time-reversal and
complex conjugation, 
i.e. (see notations above), 
\begin{align}
  \vx = \vS(\cc{\vx})_{-},\label{eq:conjsymprop}
\end{align}
then one calls $\vx$ \emph{conjugate-symmetric} and its Fourier
transform is real, i.e., we have $\FmatrixN \vx\in\R^N$. 
In \cite[Thm.2]{WJ14b} we have shown  that every signal $\vh\in\C^L$ can be conjugate-symmetrized in
$N=2L+1+K$ dimensions by 
\begin{align}
  \SymK(\vh) =\left( \begin{smallmatrix} 0 \\ \vh \\ \zero_{K} \\ \cc{\vh}_{-}\end{smallmatrix}\right)\label{eq:symK}
\end{align}
for any $K\in\N$. Due to the leading zero, we will in this paper  denote the symmetrization operation by
$\SymK$.
Moreover,  the chain ``symmetrization, Fourier transform and
absolute-square'', i.e, the map $|\Fmatrix\SymK|^2 \colon \C^L \to
\R^{N}$  corresponds to the Fourier transform of the circular auto-convolution of
$\vx=\SymK(\vh)$ for which it holds 
\begin{align}
   \vx_1\circledast \vx_1 - \vx_2 \circledast \vx_2 = (\vx_1-\vx_2)\circledast (\vx_1+\vx_2)
\end{align}
Note, this only holds for the auto-convolution and {\em not for the auto-correlation}. In fact, the auto-convolution equals the
auto-correlation if and only if the conjugate-symmetry property \eqref{eq:conjsymprop} holds. 

\paragraph{Stability Issues for Auto--Convolutions}

The inverse problem of determining a function on the interval from its auto-convolution is also known as the
``auto-convolution problem'' and has been investigated in the literature rarely. Its ill-posedness has been analyzed for
example in \cite{Gor94}.  
Using the auto-convolution on conjugate-symmetric signals such that $K\geq 2L+1$, then
the $K$ additional zeroes in \eqref{eq:symK}  allow us to write the circular auto-convolution as a linear
auto-convolution for which we could show in \cite{WJP15} and \cite{JW14b} a stability result, where the stability 
constant $\alp=\alp(2S)$ depends only on the support length of the signal:
\begin{align}
  \alp \Norm{\vh_1-\vh_2}_2\Norm{\vh_2+\vh_1}_2\leq
  \Norm{|\Fmatrix\SymK(\vh_1)|^2-|\Fmatrix\SymK(\vh_2)|^2}_2\label{eq:injecstabconst} 
\end{align}
for $\vh_1,\vh_2\in \Sigma_S^L$ ($S-$sparse vectors in $\C^L$).
Although the conjugate-symmetric property seems to be a difficult
constraint in certain applications, it can be very easily
constructed in a wireless communication scenario as we will 
demonstrate below.

\section{Semi-Blind Channel Estimation}
%
In an OFDM--like system the user data payload is encoded, modulated
and framed as sequences of complex data symbols $\hvs\in\C^N$ and then
transmitted by performing multiplexing in frequency (and time). Although our approach can even be formulated in a quite general
context of (pulse--shaped) multicarrier schemes, we will base our
exposition here on the simple setup of OFDM using a cyclic prefix with
$N$ tones (subcarriers). We assume (and in praxis this is always the case)
that $N$ is an \emph{even} number, given by powers of $2$. 
Transforming the OFDM symbol back to time (recall that we use
$\hat{\cdot}$ to denote Fourier transforms) and adding a cyclic prefix $CP$ of length $L$ to $\vs$ one
obtains at the receiver the noisy signal %
\begin{align}
  \vr_{N+2L-1}= \vh * \left(\begin{smallmatrix} CP\\ \vs \end{smallmatrix} \right)+ \vn_{N+2L-1}\label{eq:linconv}
\end{align}
where $\vh\in\C^L$ is the channel's impulse response of length $L$ and
$\vn_{N+2L-1}\in\C^{N+2L-1}$ denotes additive noise.  
The samples for $k\in\{L,\dots, L+N\}$ in \eqref{eq:linconv} 
represent then the \emph{circular} convolution as:
\begin{align}   
  {\vr}= {\left(\begin{smallmatrix}\vh\\\zero_{N-L} \end{smallmatrix}\right)}\circledast\vs + \vn. \label{eq:circconv}
\end{align}

\subsection{Conventional Pilot-aided Channel Estimation} \newcommand{\Dset}{\mathcal{D}} \newcommand{\Tset}{\mathcal{T}}
For the demodulation of the received symbols the channel coefficients have to be estimated. A well--investigated subject
here is pilot--aided channel estimation, where the carrier positions $\Pset=\set{Dk}{k\in [P]}$ with $PD=N$ are used for
pilot symbols $\hvu\in\C^{P}$. It is clear that this corresponds to an
ideal setup, since for real applications oversampling is used and only
tones in the middle half of the spectrum can be allocated. Several
approaches are known in literature to cope with this issue.
To hold our
exposition of the idea simple, we will ignore this here.
On the remaining subcarriers $\Dset\subseteq [N]\setminus\Pset$ the
user-data payload is modulated.
Writing \eqref{eq:circconv}  in the Fourier domain gives therefore on the pilot tones 
\newcommand{\AC}{{\vA^{\text{(c)}}}}
\newcommand{\AP}{{\vA^{\text{(p)}}}}
\begin{align}   
    \hat{\vec r}_{\Pset}  
    &=\sqrt{N}\diag(\hvu)\Big(\FmatrixN \big(\begin{smallmatrix}\vh\\ \zero_{N-L}\end{smallmatrix}\big)\Big)_{\Pset} +
    \hvn_{\Pset}\\ 
    &=\sqrt{P} \diag(\vu) \FmatrixP 
    \Big(\begin{smallmatrix}\vh\\ \zero_{P-L}\end{smallmatrix}\Big) + \hvn_{\Pset}
    =: \AC\vh+\hvn_{\Pset}
\label{eq:classicalce}
\end{align}
since the Fourier transform of the transmit signal $\hvs$ is constructed such that
$\hvs_\Pset=\hvu$. Thus, the received pilots tones contain a
\emph{linear} measurement of the channel impulse response and the
measurement matrix is given by $\AC$.
Here we assume $P\geq 2L$ such that we can down-sample the Fourier transform $\FmatrixP$ in $P$ dimensions. 
%
\paragraph{Least--Squares Estimation:} The standard (Tikhonov regularized) least--squares estimates (LS) for $\vh\in\C^L$
given the received pilots values $\hat{\vr}_{\Pset}$ is:
\begin{equation}
  (\AC^{*} \AC+\tau)^{-1}\AC^{*}\hat{\vr}_{\Pset}
  \label{eq:classic:lsestimate}
\end{equation}
and is suitable as long as $\AC^*\AC$ is well--conditioned which depends on $P$ and $L$. The parameter $\tau$ will be chosen
depending on the noise power $\sigma^2$, in particular for $\tau=\sigma^2$ this is also known as the Gauss--Markov
estimator.   
\paragraph{Estimation with Sparsity Priors} If we additionally know the length of the support $|\supp(\vh)|\leq S\ll L$, i.e.,
$\vh\in\Sigma^L_S$ is $S-$sparse in $L-$dimensions, then we can even use ``Compressed Sensing'' methods to reconstruct $\vh$, for example, by solving
the convex \emph{basis pursuit denoising} (BPDN) problem
\begin{equation}
  \min_{\vh\in\C^{L}} \lVert \vh\rVert_{1}\quad\text{s.t.}\quad \lVert
  \AC \vh-\hat{\vec r}_{\Pset}\rVert_{2}\leq \epsilon
  \label{eq:classic:bpdn}
\end{equation}
which yields a  stable and robust solution if $\AC\in\C^{N\times L}$
fulfils the RIP condition of order $2k$ with sufficiently small
RIP--constant (can be achieved w.h.p. by randomizing the pilot
positions), see \cite{Can08} and \cite{BHSN10}.  
Here $\eps^2$ is usually
proportional to the noise power $\sigma^2$ and has to be tuned
depending on several system parameters. In real--world wireless
applications this problem has to be solved by efficient greedy methods.

\subsection{Phaseless Channel Estimation}

\begin{figure}[ht]
    \includegraphics[width=0.98\linewidth]{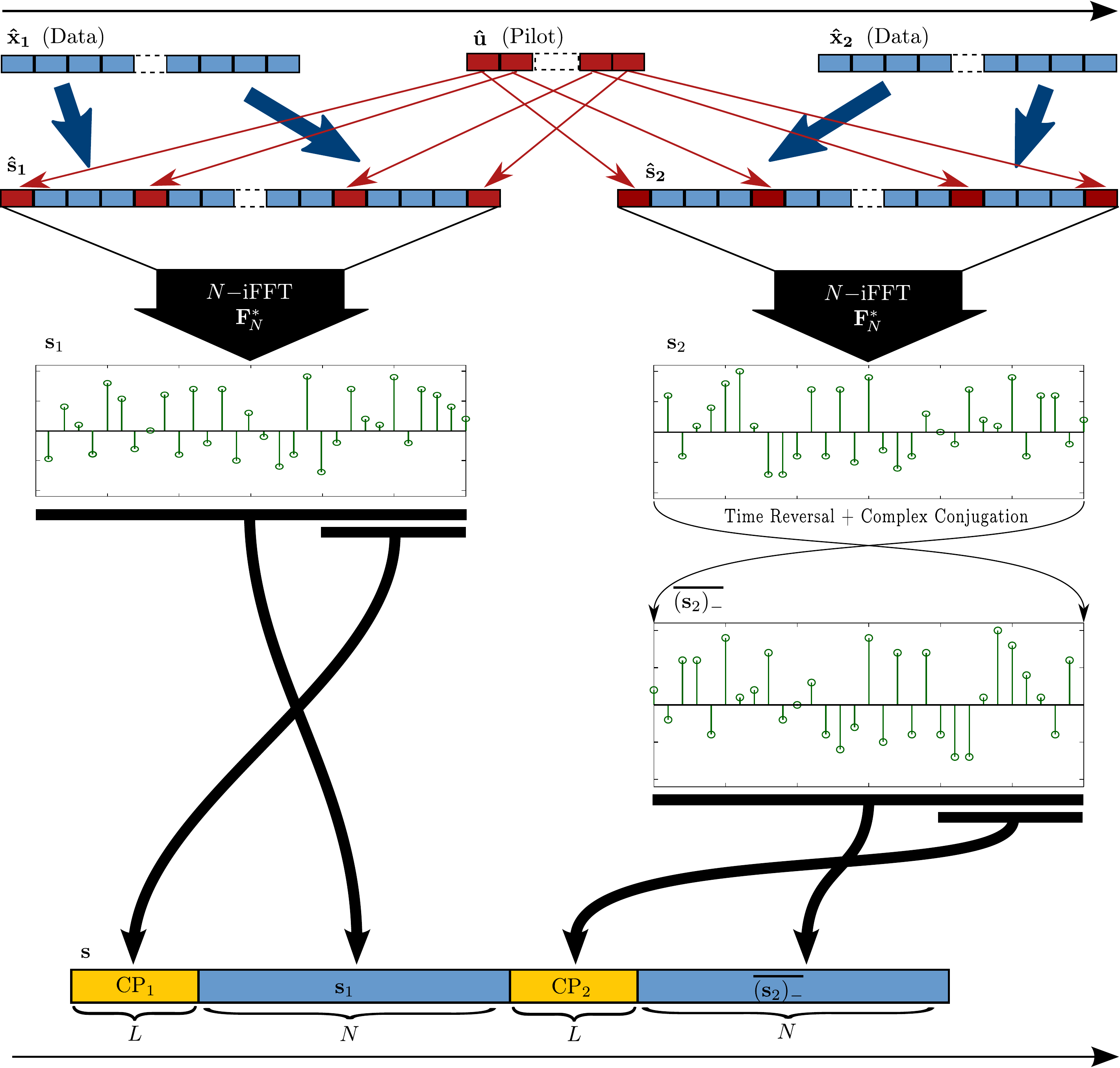}
  \caption{Scheme of the OFDM transmitter using phaseless pilots}\label{fig:sendingscheme}
\end{figure}

In our proposed scheme now, we consider \emph{two OFDM symbols} $\hvs_1$ and $\hvs_2$ carrying the same pilot tones
$\hvu$ and being transmitted over the same channel $\vh\in\C^L$.
In the shortest setup both symbols are sent consecutively.
The key feature of our scheme is to use in the
second symbol the same pilots $\hvu$ and send them time reversed and complex conjugated $(\cc{\vs_2})_{-}$ over the
channel in time, see also \figref{fig:sendingscheme}, as
\begin{align}
  \vh*
  \left[\left(\begin{smallmatrix} 0\phantom{_1}\\CP1\\ \vs_1 \\ \zero_L \\ \zero_{N} \end{smallmatrix}\right)
  +\left(\begin{smallmatrix}  0\phantom{_L} \\ \zero_L\\ \zero_{N} \\ CP2 \\(\cc{\vs_2})_{-}\end{smallmatrix}\right)\right]
        =\left(\begin{smallmatrix} \vg_1 \\ \vy_1 \\ \vg_2+\vg_3 \\ \vy_2 \\ \vg_4 \end{smallmatrix}
          \right)\in\C^{2N+3L}\label{eq:noiselessreceive2}
\end{align}
where $\{\vg_i\}_{i=1}^4\subset\C^L$ are ``garbage'' vectors due to the cyclic prefix. Hence in the noiseless case we
receive by \eqref{eq:noiselessreceive2} 
\begin{align}
    \vy_1 = \left(\begin{smallmatrix}\vh\\\zero_{N-L}\end{smallmatrix}\right)\circledast \vs_1 \quad\text{and}\quad
      \vy_2 = \left(\begin{smallmatrix}\vh\\ \zero_{N-L}\end{smallmatrix}\right) \circledast (\overline{\vs_2})_-
      \label{eq:y1y22}.
\end{align}

\noi By applying complex-conjugation and time-reversal on $\vy_2$ and shifting afterwards circular by $-1$, we can reverse the
flipping on $\vs_2$ and flip instead the channel:
\begin{align}
  \vS^{-1}(\cc{\vy_2})_{-} 
  &= \vS^{-1} \left(\left(\begin{smallmatrix}\cc{\vh}\\ \zero_{\!N\!-\!L}\end{smallmatrix}\right) \circledast
  ({\vs_2})_-\right)_{-}
  =\left( \begin{smallmatrix}\zero_{\!N\!-\!L\!}\\ \cc{\vh}_{-}\end{smallmatrix}\right) \circledast \vs_2 
\end{align}
If we take the DFT of the sum of $\vS\vy_1$  and $\vS^{-1}(\cc{\vy_2})_-$ we get:
\begin{align}
  \hvy&=\left(\Fmatrix\vS\vy_1 + \Fmatrix\vS^{-1}(\cc{\vy_2})_{-}\right)_{\Pset} \\
   &= \sqrt{N}\left[ 
      \left(\Fmatrix \left(\begin{smallmatrix}  0 \\ \vh \\ \zero_{\!N\!-\!L\!-\!1} \end{smallmatrix}\right)\right) \odot
    \hvs_1
    + \left(\Fmatrix\left(\begin{smallmatrix} 0 \\ \zero_{\!N\!-\!L\!-\!1} \\ \cc{\vh}_{-}
    \end{smallmatrix}\right)\right) \odot
    \hvs_2
  \right]_{\Pset}. \notag
\intertext{%
Since it holds by construction of the symbols $(\hvs_1)_{\Pset} = (\hvs_2)_{\Pset}=\hvu$ we can  factorize the
pilots out by bilinearity of the convolution and get}
 \hvy &= 
  \sqrt N\left( \vec F
  \left(\begin{smallmatrix}  0\\ \vh \\ \zero_{K} \\ \vhtr \\  \end{smallmatrix}\right)\right)_{\Pset} \odot
  \hvu  = \sqrt{{N}}\FmatrixN  \SymK(\vh) \odot \hvu,\label{eq:psetfouriersum}
\end{align}
with $K=N-2L-1$. By assumption $N$ is even and hence $K$ is always
odd. Hence we obtained the Fourier transform of the conjugate-symmetrized form $\SymK(\vh)$ of the impulse response $\vh$  in $N$ dimensions. Since the $P$ pilots are
uniformly spaced in $[N]$ with distance $D$ we have $N=DP$. 
If we also assume $P$ to be {\bfseries even} (we could also
consider $P$ to be odd, but then we need another symmetrization) we only need 
$K'=P-2L-1$ odd zeros between $\vh$ and $\cc{\vh}_-$ by  applying, as in the classical case, the uniform circular
sampling theorem (Nyquist Sampling)
\begin{align}
  \left(\FmatrixN \SymK(\vh)\right)_{\Pset}
  &= \sqrt{\frac{P}{N}}\FmatrixP  \SymtK(\vh)= \frac{1}{\sqrt{D}}\FmatrixP  \SymtK(\vh)
\end{align}
which is again the Fourier transform of a circular conjugate-symmetric vector, but now in  only $P$ dimensions, since
the Fourier measurements in \eqref{eq:psetfouriersum} are actually an oversampling by the factor $D$. 
Hence, taking the square of the absolute-values in
\eqref{eq:psetfouriersum} gives our noise-free \emph{processed
  Fourier measurements}
%
\begin{align}
  \vz=\frac{\diag(|\hvu|^{-2})}{P} \abs{ \hvy}^2= \left|\FmatrixP \SymtK(\vh)\right|^2
    \label{eq:receivedpilotsunshift}
\end{align}
which constitute therefore a \emph{phase retrieval problem from quadratic
measurements}: 
\begin{flalign}
 (P_{\text{quad}})&& \text{find } {\vh\in\C^{L}} \text{ s.t. } \vz=|\FmatrixP \SymtK(\vh)|^2.  &&
\end{flalign} 
By \eqref{eq:injecstabconst}, shown in \cite{WJ14b}, we can recover in
the noise-free case $\vh$ from the 
\emph{processed Fourier measurements}
$\vz$ up to a global sign, i.e., the problem $(P_{\text{quad}})$ has a unique solution up to global sign.
This should be compared to ``pure'' Fourier measurements where recovery
is up to several ambiguities, as already mentioned above (or see again \cite{BP15}).
The main idea here is to use the real-valued property of the symmetrized channel $\SymK(\vh)$. Note, that $\hvy$ is not
real valued, since the pilots $\hvu$ are not real valued. Hence taking the absolute value eliminates the unknown pilot
phases but keeps all available channel information at the receiver
except for a global sign. 

Before proceeding to reconstruction algorithms we shall elaborate more
on symmetrization as a real--valued mapping.
We will rewrite the symmetrization operator $\SymtK$, which is a \emph{``non-complex-linear''} map from $\C^L$ to
$\C^P$ as a linear map $\vLam\colon\R^{2L} \to \C^{P}$. 
Let be
$\vg=\left(\begin{smallmatrix}\vc\\\vd\end{smallmatrix}\right)\in\R^{2L}$
where $\vc=\Re(\vh)\in\R^L$ and $\vd=\Im(\vh)\in\R^L$. Then:
\begin{align}
  \vx=\SymtK(\vh) 
  =\left(\begin{smallmatrix}  0\\ \vc+i\vd \\ \zero_{K'} \\ \vc_- -i\vd_-   \end{smallmatrix}\right)
  = \left(\begin{smallmatrix} \zero_{1\!\times\! L} & \zero_{1\!\times\! L} \\ \id_L & i \id_L\\ \zero_{K'\!\times\! L}
    &\zero_{K'\!\times\! L} \\ \vR_L & -i\vR_L  \end{smallmatrix}\right)
  \begin{pmatrix}\vc\\\vd\end{pmatrix}=:\vLam\vg.
\end{align}
where $\vR_L$ denotes the matrix for time--reversal, see \eqref{eq:notation:shift:treverse}.
This gives the following equivalent of \eqref{eq:injecstabconst} for any $\vg_1,\vg_2\in\R^{2L}$
\begin{align}
  \alp \Norm{\vg_1-\vg_2}_2 \Norm{\vg_1+\vg_2}_2 \leq \Norm{|\widehat{\vLam\vg_1}|^2 -|\widehat{\vLam\vg_2}|^2}_2.
\end{align}
We can even get rid of the squares in our absolute values by  using the binomial formula by setting
$\hvx_i=\widehat{\vLam\vg_i}$ for $i=1,2$.
%
\begin{equation}
  \begin{split}
    \lVert|\hvx_1|^2& -|\hvx_2|^2\rVert_2
    =
    \Norm{(|\hvx_1| -|\hvx_2|)\odot(|\hvx_1|+|\hvx_2|)}_2\\
    &=
    \Norm{(|\hvx_1| -|\hvx_2|)^2\odot(|\hvx_1|+|\hvx_2|)^2}^{\frac{1}{2}}_1\\
    &\leq \left(\lVert(|\hvx_1| -|\hvx_2|)^2\rVert_2\cdot \lVert(|\hvx_1|+|\hvx_2|)^2\rVert_2\right)^{\frac{1}{2}}\\
    & =\Norm{|\hvx_1| -|\hvx_2|}_4 \cdot \Norm{|\hvx_1|+ |\hvx_2|}_4 \\ 
    &\leq \Norm{|\hvx_1|-|\hvx_2|}_4 \big(\Norm{|\hvx_1|^2}_2^\frac{1}{2} +\Norm{|\hvx_2|^2}_2^\frac{1}{2}\big). \\
  \end{split}
\end{equation}
Due to the conjugate-symmetry and the zero-padding in $\vx_i$ we
can upper bound (using Parseval--identity, Young and Cauchy--Schwartz inequality) to get:
$\Norm{|\hvx_i|^2}_2=\Norm{\vLam\vg_i\circledast\vLam\vg_i}_2=\Norm{\vLam\vg_i*\vLam\vg_i}_2\leq
\sqrt{2L}\Norm{\vg_i}_2^2=\sqrt{2L}\Norm{\vh_i}_2^2$. 
Assuming an upper bound $\Norm{\vh_i}_2\leq\eta$ gives therefore
the weaker stability constant: 
\begin{align}
  \alp \cdot \frac{\Norm{\vg_1-\vg_2} \Norm{\vg_1+\vg_2}}{2(2L)^\frac{1}{4}\eta}\leq
  \Norm{|\widehat{\vLam\vg_1}|-|\widehat{\vLam\vg_2}|}_4.
\end{align}
Thus, increasing the power $\eta$ decreases the stability of the map $|\widehat{\vLam\vg}|$.  We will see in the
next paragraph, that in the noisy case that this has impact on
$(P_{\text{quad}})$. However, in the noiseless case the problem based
on (non--squared) absolute-values in \eqref{eq:receivedpilotsunshift}
\begin{flalign}
 (P_{\text{abs}})&& \text{find } {\vg\in\R^{2L}} \text{ s.t. } \vz=|\FmatrixP \vLam\vg|.  &&
\end{flalign} 
is equivalent to $(P_{\text{quad}})$.
Moreover, $(P_{\text{abs}})$  constitute a \emph{generalized phase
retrieval problem for real-valued vectors}. 
\paragraph{The Noisy Versions of $(P_{\text{quad}})$ and $(P_{\text{abs}})$}
Applying additive noise to the \emph{linear} Fourier measurements 
\emph{prior} to taking magnitudes comes therefore
with a noise model that is usually \emph{not} investigated by the phase retrieval community.
Let us consider additive complex-valued noise $\vn_1,\vn_2\in\C^N$
in \eqref{eq:circconv} affecting channel estimation and data demodulation. Let us set $\valp=|\hvu|^{-1}/\sqrt{P}$ and compute
for the absolute-valued measurements
%
\begin{align}
  \vb&=\,\valp\odot\left|\Fmatrix[\vS(\vy_1+\vn_1) +\vS^{-1}(\cc{\vy_2+\vn_2})_{-}]\right|_{\Pset}   \\
     &=: \valp\odot\Big|\underbrace{(\Fmatrix[\vS(\vy_1) +\vS^{-1}\cc{\vy_2}_{-}])_{\Pset}}_{=\hvy}+\hat{\vn}\Big|.\label{eq:vbnoise}
\end{align}
Absolute-squared measurements contain therefore a cross term
$\valp\odot (\hvy\odot \cc{\hvn} + \cc{\hvy}\cdot \hvn)$ between noise
and the symmetrized channel. The
residual is therefore $\vb^2 - \valp^2\odot |\hvy|^2= \valp^2\odot(\hvy \odot \cc{\hvn} + \cc{\hvy}\odot \hvn + |\hvn|^2)$ 
and increasing SNR increases the residual as well. 
Therefore, careful adjustment of an effective (practically reasonable)
bound $\eps_{\text{eff}}$ on the expected residuals depending
apriori--knowledge or on the
statistic of system parameters is mandatory for the \emph{quadratic} problem:
\begin{flalign}
  (P_{\text{quad},\epsilon_{\text{eff}}})&& \text{find } {\vg\in\R^{2L}} \text{ s.t. } \Norm{\vb^2-|\FmatrixP \vLam\vg|^2}_2 \leq \epsilon_{\text{eff}}&&
\end{flalign} 
For the (non--squared) absolute-valued problem $(P_{\text{abs}})$ we
compute the residual between $\vb$ in \eqref{eq:vbnoise}
and the noise-free version $\vz$ in \eqref{eq:receivedpilotsunshift}
using reverse triangle inequality:
%
%
\begin{align}
  \Norm{\vb - \vz}_2=\Norm{\valp\odot\left(|\hvy+\hvn| - |\hvy|\,\right)}_2 \leq \norm{\valp\odot \hvn}_2.
\end{align}
%
%
From our stability result and the fact that $f(\vg)=|\widehat{\vLam\vg}|=\vz$ is convex we have the following convex optimization
problem for the \emph{noisy phase retrieval problem for absolute Fourier measurements of symmetrized signals}: 
\begin{flalign}
  (P_{\text{abs},\eps})&& \text{find } {\vg\in\R^{2L}} \text{ s.t. } \Norm{\vb-|\FmatrixP \vLam\vg|}_2 \leq \eps &&
\end{flalign} 
where we simplified by $|\alp_k|=1$ and $\Norm{\hvn}_2\leq \eps$. 
Obviously, this problem is still ill-posed. We will postpone the
investigation of $(P_{\text{abs},\eps})$ to a companion paper.
The problem can not be solved directly using an linear
estimator as compared to $(P_{\text{quad},\epsilon_{\text{eff}}})$
(without further assumptions).

\section{Algorithms}

For the quadratic problem $(P_{\text{abs},\eps})$ we  propose here two
different reconstruction strategies. First, we tackle the problem in a direct way by lifting the quadratic problem to a linear
low-rank matrix recovery problem. Its convex relaxation can be efficiently
solved by a semidefinite program.  This approach follows the same
lifting ideas recently used in phase retrieval by \cite{CSV12} but
with specific (not Gaussian, not generic and even not Fourier),
structured measurements. Although this is motivated by the
noise--robustness of such programs its practical implementation is at
this time not yet feasible for the desired application.

The second strategy is a two-stage approach, where we first solve the 
linear problem of estimating the auto-convolution of the symmetrized
channel and afterwards performing an iterative deconvolution algorithm
to 
extract the channel. Nevertheless, the iterative ``de-autoconvolution'' has also its stability issue and is in general
difficult to address, see e.g. \cite{Gor94}.

\subsection{Recovery via a Semidefinite Program}
%
For any $\vx\in\C^P$ we can write $|\FmatrixP\vx|^2$ as a linear map of a rank$-1$ matrix $\vx\vx^*$, which is known as the (phase)
lifting technique \cite{CSV12},\cite{CESV13}. For each $k\in[P]$ we have:
\begin{align}
  \left(\left|\FmatrixP \vx \right|^2\right)_k
  =\tr(\ihve_k\ihve_k^*\,\vx\vx^*)\label{eq:phaseliftreal}
\end{align}
and overall this maps complex-valued positive matrices  to positive real-valued vectors.  Incorporating
the conjugate-symmetric structure $\vx=\vLam\vg$  in \eqref{eq:phaseliftreal} yields for $k\in[P]$
\begin{align}
   z_k=|\FmatrixP\vLam\vg|^2_k =\tr(\left(\vLam^*\ihve_k\ihve_k^*\vLam\right)   \vg\vg^T)=:\A(\vg\vg^T)_k\label{eq:liftedtrace}.
\end{align}
where by the conjugate-symmetry property \eqref{eq:conjsymprop} this defines indeed a real-valued linear map
$\A:\R^{2L\times 2L}\to \R^{P}$.  This linear map can also be seen as the square of the superposition of an oversampled
discrete cosine transform on the real part and an oversampled discrete sine transform on the imaginary part of $\vh$.
Similar approaches are used also in \cite{FMNW14,BH13,Phi14}.  From \eqref{eq:injecstabconst} we know that only
one positive rank$-1$ matrix $\vG=\vg\vg^T\in\R^{2L\times 2L}$ exists giving noise-free
measurements $\vz$ in \eqref{eq:receivedpilotsunshift}. The minimal amount of any real-valued linear measurements to
guarantee unique reconstruction up to global sign is $2(2L)-1=4L-1$, which can be achieved by generic linear
measurements \cite{BCE06}.  In our case we found  $P\geq 4L+2$ deterministic real-valued linear measurements to
guarantee recovery up to global sign. In fact, the leading zero for the channel symmetrization could even be omitted by
assuming $h_0$ to be real, which would give unique recovery of $2L-1$ real-valued unknown from only $4L-2$ magnitudes of
linear measurements.  Hence we end up with the following  rank minimizing optimization  
\begin{flalign}
  (P_0)&& \vG=\argmin_{\vtG\in\R^{2L\times 2L}, \vtG\geq \zero} \rk(\vtG)  \quad \text{s.t.}\quad  \vz=\A(\vtG)&&
  \label{eq:rankmin}
\end{flalign}
which is equivalent to $(P_{\text{quad}})$. The real vector $\vg$ is
obtained from the SVD of $\vG$  and hence
$\vh=(g_0+ig_{L},\dots,g_{L-1}+ig_{2L-1})^T$ up to global sign. 
To determine the global sign it is sufficient to know one of the phases of
the pilot tones or one can even reconstruct this from the data under
certain assumptions.
The following convex relaxation of \eqref{eq:rankmin} is well--established in
phase retrieval \cite{CZ13},\cite{RFP10}, \cite{CESV13}:
\begin{flalign}
  (P_*)&& \vG=\argmin_{\vtG\in\R^{2L\times 2L}, \vtG\geq \zero} \trace(\vtG)  \quad \text{s.t.}\quad  \vz=\A(\vtG)&&
  \label{eq:tracemin}
\end{flalign}
In fact, one still has to show that all other semidefinite-positive matrices which are feasible solutions
$\set{\vtG\in\R^{2L\times 2L}}{\A(\vtG)=\vz}$ yielding strictly larger traces. To derive such results
further assumptions on the signals are needed \cite{JEH15}. At time of
writing it is not clear to what extent the conjugate-symmetric property is
sufficient here. In our simulation we observed that there might exists other matrices of higher rank
yielding smaller traces than $\vg\vg^T$, although for highly sparse channels $(P_*)$ seems to yield
satisfying results. In this case we can promote sparse matrices in $(P_*)$ by the convex function
$\lam\Norm{\vG}_1$ for some parameter $\lam>0$:
\begin{flalign}
  (P_{*,\eps_{\text{eff}}})&& \min_{\substack{\vtG\in\R^{2L\times 2L}\\  \vtG\geq \zero}} \tr(\vtG)
  +\lam\Norm{\vtG}_1 \,\, \text{s.t.}\,\,
  \Norm{\vz-\A(\vtG)}_2\leq\eps_{\text{eff}} &&\notag
\end{flalign}
where the discussion on the effective bound $\eps_{\text{eff}}$ above
is relevant again, i.e., $\eps_{\text{eff}}$ has to be manually
adjusted with respect to prior-knowledge on $\vh$ or to some
statistical parameters.  
The problem above is a semidefinite program and it can be solved using,
for example, Sedumi. On the other, from practical perspectives such
algorithms are far from being usable in the context of channel
estimation due to complexity reasons.
Furthermore, at time of writing, this does not provide a sufficiently robust
channel estimation for practical relevant SNR values which is caused
by the noise cross--term issues described above. Therefore we propose an iterative-deconvolution
algorithm, which indeed produces similar BER rates for the channel estimation as in the classical cases with known pilot
phases, see \figref{fig:ber}.

\subsection{Two--Stage Iterative Recovery}
The absolute-square  Fourier measurements of $\SymtK(\vh)$ can also be seen as the DFT of the circular auto-correlation
of $\SymtK(\vh)$. Due to the conjugate-symmetry this equals the circular auto-convolution with 
\begin{align}
  \SymtK(\vh)\circledast \SymtK(\vh)
  =\vS^{-2L}\left(\begin{smallmatrix} \va \\   \zero_{P-4L-1}\end{smallmatrix}\right)
\end{align}
with $\va=\Symnull(\vh)*\Symnull(\vh)\in\C^{4L+1}$ since $P\geq 4L+1$ and $\vh\in\C^L$, see also \cite{WBJ15a} where we
also discuss the case of non-uniform pilots, due to dimensions mismatches.
Due to the conjugate--symmetry of $\va$ it is also sufficient to
estimate only the one--sided autoconvolution, i.e., optimize in the
first stage over vectors in $\C^M$ where $M=2L+1$ instead of $4L+1$.

\paragraph{Least--Squares Estimation of the Auto-convolution}
Hence, the first stage is a least-square (LS) estimation of  the auto-convolution $\va$ from the square of the noisy measurements $\vb$ in
\eqref{eq:vbnoise} given by
\begin{align}
  \min_{\vta\in\C^{M}} \Norm{\vb^2 - \AP\vta}
\end{align}
with the linear map $\AP=\Fmatrix \vS^{-2L}\vQ\in\C^{P\times M}$
where $\vQ\vx$ 
zero-pads $\vta$ to $P$ dimension. Similar
to \eqref{eq:classic:lsestimate} we can  consider here the family:
\begin{equation}
  (\AP^{*} \AP+\tau)^{-1}\AP^{*}\vb^2
  \label{eq:phaseless:lsestimate}
\end{equation}
of estimates for depending on a choosen parameter $\tau$.

\paragraph{Estimation of the Auto-convolution under Sparsity Prior}

If the channel impulse response is known to be $S$--sparse its symmetrized version is at most $2S-$sparse and hence its
auto-convolution is at most $(2S^2+S)$--sparse.
\begin{equation}
  \min_{\vta\in\C^{M}} \Norm{\vta}_0 \quad\text{s.t.}\quad \Norm{\vb^2- \AP\vta}_2\leq \epsilon_{\text{eff}}
  \label{eq:classic:bpdn2}.
\end{equation}
If $\AP$ acts almost isometric on the desired $\vta$'s
the problem above can be relaxed \eqref{eq:classic:bpdn2} to
the (convex)
\emph{basis pursuit denoising problem}
\begin{align}
  \min_{\vta\in\C^{4L+1}} \Norm{\vta}_1 \quad\text{s.t.}\quad \Norm{\vb^2-\AP\vta}_2\leq \eps_{\text{eff}} \label{eq:csphaseless}
\end{align}
But, due to the increased sparsity of the auto--convolution with
symmetrization the gains are limited here for practical
settings. As already discussed the overall compressibility
due to a sparse channel can not fully taken into account here since
the original problem is an sparse rank--one problem. Furthermore,
randomizing pilot position in a fashion known to the receiver
increases again the amount of prior--knowledge at the receiver which 
we want to avoid.

\paragraph{Iterative De-Autoconvolution, Noiseless Case} For a first exposition we start with the noiseless case and the
ideas was inspired by the work of \cite{FMNW14} and \cite{BH13}. A
similar approach has been used in \cite{Phi14}. 
However, the algorithms proposed in these works are not noise
robust. An overall stable ``de-autoconvolution'' is indeed an unsolved problem \cite{Gor94}.
Nevertheless, we present an approach of a  manually adopted  iterative Tikhonov regularization.  The values $h_0\dots
h_{L-1}$ are determined up to global sign by $a_0\dots a_{L-1}$ due to:
\begin{equation}
  \begin{split}
    a_k 
    &= \sum_{l=0}^{k} h_l h_{k-l}
  \end{split}
\label{result_correlation}.
\end{equation}
To see this, let us assume that $a_0$ is non--zero and real
(multiplying $\va$ by a global phase).  
%
%
Then it follows from \eqref{result_correlation} that
$a_{0}= h_0^2$, i.e., $h_0=\pm\sqrt{a_0}$.  Given the values $h_{0}\dots h_{k-1}$ one can can solve for $h_k$ using again
\eqref{result_correlation}:
\begin{align}
  h_k &= \frac{a_{k} - \sum_{l=1}^{k-1} h_l h_{k-l}}{2 h_0}.
\end{align}
At this point it is clear that dividing by $h_0$ could make this algorithm quite unstable in the noisy setting.  For
practical purposes it is therefore important to estimate $h_0$ with
high accuracy and to order operations by magnitudes (which we will not
consider here).

\paragraph{Noisy Case, Thresholding and Regularization}

Let $\vec X=\vh\vh^T$ be the rank--one $L\times L$ matrix associated to the vector
$\vh$. Since $\vh\in\C^L$ is complex the matrix $\vec X=(X_{ij})\in\C^{L\times L}$ is \emph{not} positive
semi--definite. Let $X_{ij}=\langle \ve_i,\vec X\ve_j\rangle=\trace(\vec X\ve_i\ve^T_j)$ be the elements
of $\vec X$ where $\ve_i$ are the real standard basis vectors. 
We can write therefore:
\begin{equation}
  a_k \overset{\eqref{result_correlation}}{=}\sum_{l=0}^k X_{l,k-l}
  =\trace\left(\vec X\sum_{l}\ve_{k-l}\ve^T_l\right)=:\trace(\vec E_k\vec X)
\end{equation}
where the matrix $\vec E_k$ shifts coordinates $1\dots k$ (in the standard basis) of a vector by $k$ positions. Consistent
with our first illustration of the noiseless case above, this means that $\trace(\vec E_k\vec X)$ depends \emph{only} on
$h_{0}\dots h_{k}$.  Although this is advantageous from complexity point of view this is also the reason for error
accumulation. On the other hand the equation above allows for post--processing the values $h_0\dots h_k$ by a
least--squares approach with regularization in the sense of Tikhonov. This regularization has already been investigated and
proposed for general autoconvolution problems, see \cite{Gor94}.  More precisely, let $\va^{\#}$ be the estimate of the
autoconvolution obtained from stage one in a noisy setup. In
applications, the channels to be estimated are often sparse or
compressible, i.e. for example $\vh$ is $S$--sparse. Thus it is reasonable to sparsify noisy measurements of its
autoconvolution by a thresholding procedure:
\begin{equation}
  \va^{\text{thr}}=\va^\#\cdot\mathbf{1}(k>0:|a_k^\#|\geq\lambda).
\end{equation}
In particular, $k=0$ will be ignored here since it will be used to determine $h_0$ (although there would be
alternatives).  Hard thresholding in the order of a fraction of the universal threshold according to \cite{DJ94} (for
Gaussian noise) would suggest a thresholding level of
\begin{equation}
  \lambda\approx 0.1\sqrt{\sigma^2_{\text{eff}}(4L+1)}.
\end{equation}
In our experiments we observed that this already gives a substantial
improvement. However, in real applications this ratio has
obviously to be tuned to the environment and to the statistics of all random contributions.

The least-squares solution
$\vh$ for $\va^{\text{thr}}$, often used due to its simplicity,
would minimize the $\ell_2$--norm of the following residual:
\begin{align}
  \va^{\text{thr}} - \sum_{k}\tr\left(\vec E_k \vh\vh^T\right)\ve_k.
\end{align}
Minimizing such an objective (locally, since it is not convex in the
vector $\vh$) results in considerable strong fluctuations of
$\lVert\vh\rVert_2$ since the auto-convolution problem itself is ill--posed.
A common approach here is the following regularization:
\begin{align}
  \min_{\vh}\,\lVert \va^{\text{thr}} - \sum_{k}\tr\left(\vec E_k \vh\vh^T\right)\ve_k\rVert^2_2+
  \alpha\lVert\vOme\vh\rVert_2^2
  \label{eq:lineartikhonov}.
\end{align}
Since the objective is non-convex it can with moderate complexity only be minimized locally (which itself can be
implemented efficiently again), i.e., a reasonable initial point is necessary. Furthermore, we use
$\vOme=\diag(e^{(L-1)\ome},\dots,e^{0\ome})$ with $\omega=4$ to perform a reweighting. This causes a decreasing
influence of ``past'' values to the current iteration $k$.  Together with the observation that $\tr(\vec E_k \vh\vh^T)$
depends only on the values $h_0\dots h_{k}$ we propose the following successive approximation algorithm (see the
algorithm below). 
First, assume that values $h_0\dots h_{k-1}$ are given.  Then the value of $h_k$
is obtained according to \eqref{result_correlation}.  We use then this $k+1$-dimensional vector containing $h_0\dots
h_k$ as the initial point for locally minimizing \eqref{eq:lineartikhonov} to get an update.

Let us call the updated values (the $\argmin$ in \eqref{eq:lineartikhonov}) again with $h_0\dots h_k$. As an
intermediate operation we apply a {\em pruning step}, i.e., we keep only the $k^\text{prun}$ values having largest
magnitude. Again this parameter has to be setup depending on the conditions. In our tests we assume that the channel
$S$--sparse (or we are interested only in the $S$ values with largest magnitude) and we have used $k^\text{prun}=3S/2$.

The procedure above is iterated up to $k=L$ and finally $\vh$ is again pruned to the $S$ entries with largest
magnitude. Using this overall iterative algorithm we observed that it possible stabilize the de-auto-convolution.

\begin{algorithm}[H]
\begin{algorithmic}
  \STATE thresholding: $\va^{\text{thr}}\gets \va^\#\cdot\mathbf{1}(k>0:|a_k^\#|\geq\lambda)$
  \STATE $\vh\gets \zero, h_0 \gets \sqrt{a^{\text{thr}}_{0}}$
\FOR{$k=1$ to $L-1$}
	\STATE \COMMENT{prediction step:}
        \STATE $h_k \gets (a_k^{\text{thr}}- \sum_{l=0}^k h_lh_{k-l} )/ (2 \cdot h_0)$
	\STATE \COMMENT{local update step:}
    \STATE $\vec h \gets \underset{\vth\in\C^L\,\text{and with initial value}\,\vh  }{\text{locally minimize}}
    \left\{\norm{\begin{pmatrix}\left\{ a_{l}^{\text{thr}} - \tr\left( \vec E_l \vth
    \vth^*\right)\right\}_{l=0}^k \\ \alpha\vOme \vth\end{pmatrix}}_2^2\right\}$
      \STATE    $\qquad\text{s.t.} \quad \tilde{h}_l = 0 \;\forall\; l \in \{k+1,\ldots,L-1\}$
      \STATE \COMMENT{pruning step:} 
      \STATE $\vec h \gets k^\text{prun}$ entries of $\vh$ with
      largest magnitude (other are $0$) 
      \ENDFOR
\label{alg:itdeconv}
\end{algorithmic}
\end{algorithm}

\subsection{Performance Evaluation}
\begin{figure}[h]
  \hspace*{-1.2em}
  \includegraphics[width=1.16\linewidth]{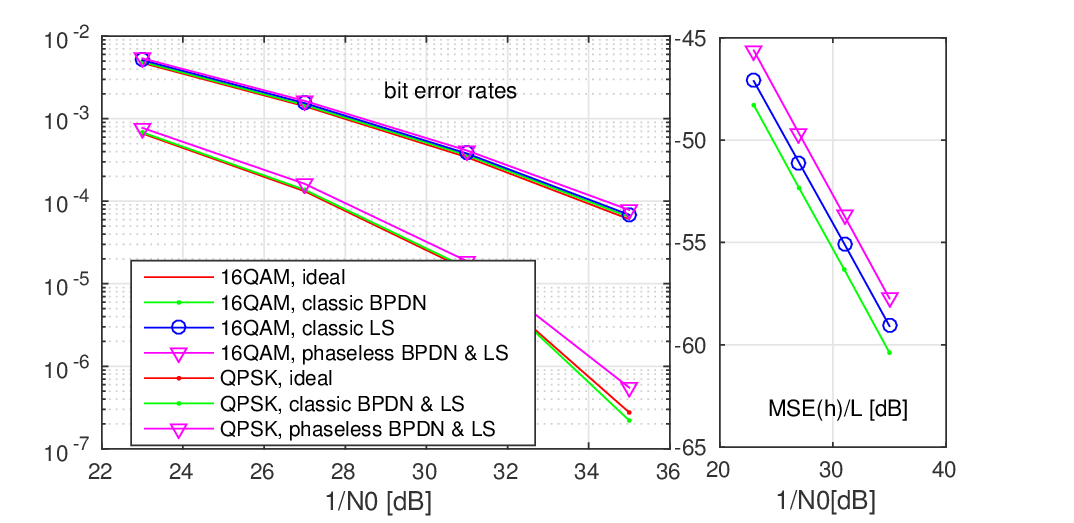}
  \caption{Bit error rate and mean squared error (MSE) per dimension for QPSK and 16QAM modulation. 
    The parameters are $N=2048$, $L=20$, $P=256$, $S=3$ and $\alpha=0.01$.
    Classical least--squares (LS) and BPDN channel estimation is
    plotted in blue and green. Red denotes ideal channel
    knowledge. The cyan curve is 
    the phaseless scheme where the autoconvolution is LS- or 
    BPDN--estimated (no difference). The x--axis is inverse power
    $N_0$ [dB], i.e., the variance of i.i.d. complex normal
    distributed noise $\hvn$.}\label{fig:ber}
\end{figure}

We have evaluated the final performance in terms of (uncoded) bit
error rates in a physical layer OFDM simulation. We observed
that under moderate system conditions similar  performance can
be obtained for phaseless estimation as compared to
conventional pilot--aided channel estimation, see Fig. \ref{fig:ber}.
On the other hand, we also note that at the time of writing the
iterative algorithm for phaseless estimation still require careful adjustment
of several parameters, as discussed above.

\section{Conclusion} 
We have presented a novel semi-blind channel estimation method in using phaseless pilots for OFDM.
With our approach the pilot phases are available again for
user--dependent information. We have discussed the relation and the differences to phase
retrieval using Fourier measurement and proposed a lifting method for
our setup. On
the numerical side we have established a two--stage
algorithm which first estimates the auto--convolution and then proceeds
by recovering iteratively the channel impulse response in a
regularized fashion.
Our results show that under moderate
system assumptions it is possible to stabilize the ``de--autoconvolution'' and
similar performance in BER can be obtained as compared to classical
channel estimation using pilot phases.

{\it Acknowledgments.} We would like to thank Holger Boche for helpful discussions. This work was
partially supported by the DFG grant JU 2795/2 and WA 3390/1.

\printbibliography

\end{document}